\documentclass[aps,preprint]{revtex4} 
\usepackage [dvips]{graphicx}   
\newcommand{\ds}{\displaystyle}
\newcommand{\be}{\begin{equation}}
\newcommand{\ee}{\end{equation}}
\newcommand{\ba}{\begin{eqnarray}}
\newcommand{\ea}{\end{eqnarray}}
\newcommand{\bea}{ $^{10}$Be }
\newcommand{\beb}{ $^{11}$Be }
\begin{document}
\title{Deformation or spherical symmetry in \bea and the inversion of
  1/2$^-$-1/2$^+$ states in \beb.}
\author{N. Vinh Mau}
\affiliation{Institut de Physique Nucl\'eaire, IN2P3-CNRS, 
   Universite Paris-Sud, F-91406 Orsay Cedex, France}  

\begin{abstract}
 For a core plus one neutron system like $^{11}$Be we have calculated the
 energies of the 1/2$^-$ and 1/2$^+$ states assuming a deformation of the
 core  deduced from the low energy 2$^+$ state
properties or taking into account  the coupling of the neutron with 
this 2$^+$ state
interpreted as a spherical one-phonon state. We have shown that the two
derivations yield identical results if  the phonon
energy is neglected in the second derivation and  close results in the general
case.   
\end{abstract} 
\maketitle
The problem of the 1/2$^-$-1/2$^+$ states inversion in \beb has been for long 
a challenge for
theoreticians and has been the subject of a large number of publications
assuming deformation \cite{nu,ra,es,li} or spherical symmetry
\cite{vm,sa,ot,des}.
However we restrict our discussion to papers by Nunes et al. \cite{nu}and
Vinh Mau \cite{vm} which are
directly concerned with the inversion problem and have proposed simple
models to examine its origin. These two theoretical papers which relate the
inversion of the 1/2$^-$-1/2$^+$ states in \beb to the  existence
of a low energy 2$^+$ state in \bea are often
considered as conflicting . Indeed the first one relies on the
assumption that this  2$^+$ state at 3.36 MeV is a rotational state implying a
deformation of the nucleus while the other assumes  it to be a one-phonon
  vibrational state therefore works with a spherical nucleus. The parameter
$\beta_2$ deduced from the measured B(E2) is in the first case interpreted
as a deformation parameter   and in the second case as a collective
transition amplitude for the vibrational phonon. Intuitively we could
already say that these two interpretations are not independent or
contradictory because
deformation comes from strong correlations between nucleons inside the
nucleus   which are taken into account implicitly when $\beta_2$ is
interpreted as a collective amplitude.  This equivalence shows up  in
large basis shell model calculations    which are able to reproduce
rotational as well as vibrational states \cite{el}. Moreover   
   a recent  analysis of p(\beb,\bea)d reaction 
\cite{pi} where the wave functions of the last neutron in \beb were taken  
from the two models leads to the same agreement with experiments.

The aim of this note is to show explicitly and analytically that there is a
simple  relation between the two derivations for 1/2$^+$ 
and 1/2$^-$ states. 
The basic assumptions of the two methods \cite{nu,vm} are schematised in Fig.1.

\begin{figure}[h]
\begin{center}
\includegraphics*[scale=0.7]{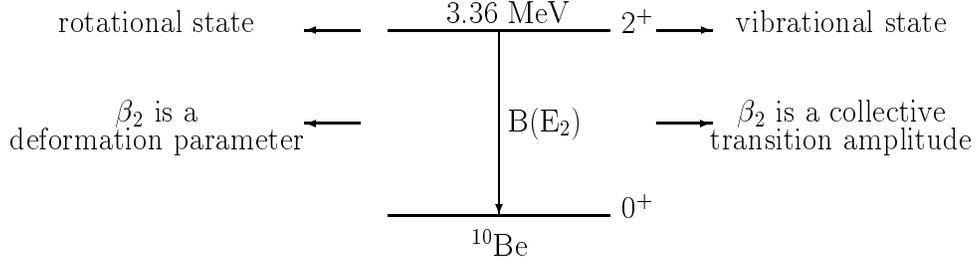}
\end{center}
\caption{Interpretation of the 2$^+$ state in $^{10}$Be as proposed in ref.[1]
(left side of the diagram) and in ref[5] ( right side of the diagram).}
\end{figure}

Let's start with the left-hand side of the figure which illustrates the
model used by Nunes et al.. The neutron one-body
potential is then written as a deformed Woods-Saxon potential by replacing
the spherical radius $R_0$ by the deformed one, $R_0(\theta,\phi)$. It writes as:
\ba
V^{def}_{nc}(r, \theta, \phi) &=&V_{0}\left(1+exp[(r-R_0(\theta, \phi))/a)
]\right)
^{-1}\\
R_0(\theta, \phi)&=&R_0\left(1+{\beta}_2Y_2^0(\theta, \phi)\right)
\ea
what, by performing an  expansion around $R_0$, leads to the potential:
\be
V_{nc}^{def}(r,\theta,\phi)\simeq V_{WS}(r,R_0)-\beta_2 R_0
{\ds\frac{dV_{WS}(r,R_0)}{dr}}Y_2^0(\theta,\phi)
\ee
The second term of eq.(3) is a correction to  the spherical Woods-Saxon
potential, $V_{WS}(r,R_0)$, and can be considered as a small 
perturbation. Therefore $\delta \epsilon_n$, the corresponding 
modification of the single neutron energy for a neutron  state
represented by the index n,  may be calculated to lowest order in perturbation
theory. For 1/2$^+$  and 1/2$^-$ states the first order term is zero and one
has to go to second order which leads to:
\be
\delta \epsilon_n=\beta_2^2 R_0^2 \sum_{\lambda\neq n} {\ds \frac{<n|
     \frac{dV_{WS}}{dr} Y_2^0|\lambda><\lambda|\frac{dV_{WS}}{dr}
      Y_2^0|n>}{\epsilon_n-\epsilon_\lambda}}
\ee
where $|n>,\epsilon_n$ and $|\lambda>,\epsilon   _\lambda$ are the
eigenvectors 
 and eigenvalues of the spherical potential, $V_{WS}$,  for states n and
$\lambda$ respectively. The summation over $\lambda$ runs over the complete
set of neutron states except state n. Performing the calculation of 
matrix elements leads to:
\be
\delta \epsilon_n=\frac{1}{10} \beta^2_2 R_0^2 \sum_{\lambda\neq n}
(-1)^{j_n-j_\lambda} {\ds
\frac{R_{n\lambda}^2}{\epsilon_n-\epsilon_\lambda}}\left|<l_n,j_n||Y_2||
l_\lambda,j_\lambda>\right|^2
\ee
where $<l_n,j_n||Y_2||l_\lambda,j_\lambda>$ is the reduced matrix element of
$Y_2$ and $R_{n \lambda}$ the radial integral:
\be
R_{n\lambda}=\int_0^\infty r^2\;dr\;\phi^*_{\alpha_n  l_n
  j_n}(r)\;\phi_{\alpha_{\lambda} l_\lambda j_{\lambda}}(r) {\ds
      \frac{dV_{WS}(r)}{dr}}
\ee
with obvious notations.

Going to the right hand side of Fig.1 which illustrates the model of ref.[5]
and after calculation of the
particle-phonon coupling  diagrams for  phonons of angular momentum L and
energy $E_L$, one
obtains  the modified one-body potential  \cite{vm} as:
\ba
V_{nc}^{sph}({\bf{r,r'}})&=&V_{WS}(r)\delta({\bf{r,r'}})+\sum_{L}{\ds 
  \frac{\beta_L^2 R_0^2}{2L+1}}\sum_{\lambda,M}
F_{L,\lambda}\phi_{\lambda}^*({\bf r})\;{\ds \frac{dV_{WS}(r)}{dr}}
Y_L^{M*}(\theta
\phi) \nonumber \\
  & &\phi_{\lambda}({\bf r}'){\ds
  \frac{dV_{WS}}{dr}}|_{r=r'}\;Y_L^M(\theta',\phi')\\
F_{L,\lambda}&=&{\ds\frac{1-n_\lambda}{\epsilon_n-\epsilon_\lambda-E_L}}+{\ds
  \frac{n_\lambda}{\epsilon_n-\epsilon_\lambda+E_L}}
\ea
where $n_\lambda$ is the occupation number of state $\lambda $  and
$ \phi_\lambda
({\bf r})$    the three dimensional wave function of the neutron calculated
in  the Woods Saxon field. 
The potential is non local, has spherical symmetry and looks very different of
the deformed potential of eq.(3). However it is again a perturbation to 
$V_{WS}$ and
to first order gives for the contribution of a 2$^+$ phonon as:
\ba
\delta \epsilon_n&= &\frac{1}{5} \beta_2^2 R_0^2 \sum_{\lambda,M}
F_{2,\lambda}<n|\frac{dV_{WS}}{dr}Y_2^M|\lambda><\lambda|\frac{dV_{WS}}{dr}
Y_2^{M*}|n> \\
&=&\frac{1}{10} \beta_2^2 R_0^2 \sum _{\lambda  \neq n}
(-1)^{j_n-j_\lambda} (\frac{1-n_\lambda}{\epsilon_n-\epsilon_\lambda-E_2}}+{\ds
 {\ds \frac{n_\lambda}{\epsilon_n-\epsilon_\lambda+E_2}}) 
|R_{n\lambda}|^2\left|<l_n,j_n||Y_2||l_\lambda,j_\lambda>\right|^2
\ea
where the term $\lambda=n$ is automatically eliminated for $j_n=1/2$
because of angular momentum coupling.
By comparing the two formulae, eqs.(5) and (10), one sees that 
they are identical
if one takes $E_2$, the phonon energy, equal to zero. Because $E_2$ appears
in the denominators only, it is easy to see that eq.(10) yields a 
correction which is smaller than given by eq.(5). However  the phonon 
energy is small (few MeV) and in the limit of very strong collectivity is
close to zero then it 
 should not introduce a too large difference between the two 
derivations.  
This result tells us  that one may not consider the two models as 
contradictory and that it is not justified to reject one or the other
as it is sometimes done.  

For other states with $j_n \neq 1/2$ we have not found such a simple relation
between the two derivations because the perturbative first order
contribution to $\delta\epsilon_n$ is non zero for the potential of eq.(3).

The theoretical expressions of the potentials have been used in different
ways. In ref.[1] $\beta _2$ was taken from the measured value of the B(E2)
and the strengths of Woods-Saxon and spin-orbit potentials fitted to the
experimental neutron energies in $^{11}$Be. In the papers following ref.[5]
the Woods-Saxon and spin-orbit potentials were fixed accordingly to their
known properties in normal nuclei \cite{bm} and the strength of the second
term of
eq.(7) was parametrised assuming a surface form factor,
${\ds{(\frac{dV_{WS}(r)}{dr})^2}}$, as suggested by
theory. When applied to reaction problems involving $^{11}$Be, 
the two potentials give the same
(good) results \cite{pi}  not only for 1/2$^+$ and 1/2$^-$ but also for
5/2$^+$ states.

We have shown that assuming a deformed mean field model or taking account of
two-body correlations in a spherical model leads to close results for the
inversion of 1/2$^+$ and 1/2$^-$ in core plus one neutron systems. We
think that this equivalence between the two models is also visible, 
 at least qualitatively, in 
the work of Li and Heenen
\cite{li}. In a deformed Hartree-Fock calculation they get the energy
minimum in \bea and \beb for a spherically symmetric configuration 
and a 1/2$^-$ ground state in \beb. The same result comes out from a
Hartree-Fock-Bogolioubov calculation \cite{gi}. The authors of ref.[4]
introduce two-body
correlations  by angular momentum projection of Hartree-Fock
wave functions. The
effect of projection on \beb is not enough to bring the 1/2$^+$ state below the
1/2$^-$ state but reduces the energy difference between them. 
The minimum of the projected \bea energy is then obtained
for a deformed configuration . We think that
this work is another proof of the close 
equivalence between the models of ref.[1]
and [5] and a justification of both: in a mean field approximation  \bea
and \beb are
spherical (then justifying the starting assumption of ref.[5]) but when one
adds correlations (what is analogous to the treatment of ref.[5]) one improves
the results in \beb and one gets a deformed configuration 
(what is a justification of the treatment of ref.[1]).

\end{document}